\input epsf
\documentstyle[12pt]{article}
        {\newpage
        \setcounter{page}{1}}

\renewcommand{\title}[1]{%
        {\begin{center}
        \Large\bf #1
        \end{center}}
        \vskip .3in}

\renewcommand{\author}[1]{%
        {\begin{center}
        #1
        \end{center}}}

\renewcommand{\abstract}[1]{%
        \begin{center}%
        {\vspace{1em}\vspace{0pt}\bf Abstract}%
        \end{center}%
        \noindent #1}

\renewcommand{\date}[1]{%
        \begin{center}%
        #1%
        \end{center}}

        {\end{thebibliography}}

\makeatletter
        \@addtoreset{equation}{section}%
\makeatother

\newcommand{\eqn}[1]{\label{eq:#1}}

\newcommand{\refeq}[1]{(\ref{eq:#1})}

\newcommand{\Eq}{Eq.~\refeq}

\newcommand{\beq}{\begin{eqnarray}}
\newcommand{\eeq}{\end{eqnarray}}

\newcommand{\eg}{{\it e.g.}}

\newcommand{\mybar}[1]%
        {\kern 0.8pt\overline{\kern -0.8pt#1\kern -0.8pt}\kern 0.8pt}
\newcommand{\sla}[1]%
        {\raise.15ex\hbox{$/$}\kern-.57em #1}
\newcommand{\roughly}[1]%
        {\mathrel{\raise.3ex\hbox{$#1$\kern-.75em\lower1ex\hbox{$\sim$}}}}

\newcommand{\drawsquare}[2]{\hbox{%
\rule{#2pt}{#1pt}\hskip-#2pt
\rule{#1pt}{#2pt}\hskip-#1pt
\rule[#1pt]{#1pt}{#2pt}}\rule[#1pt]{#2pt}{#2pt}\hskip-#2pt
\rule{#2pt}{#1pt}}

\newcommand{\Yfund}{\raisebox{-.5pt}{\drawsquare{6.5}{0.4}}}
\newcommand{\Ysymm}{\raisebox{-.5pt}{\drawsquare{6.5}{0.4}}\hskip-0.4pt%
        \raisebox{-.5pt}{\drawsquare{6.5}{0.4}}}
\newcommand{\Ythrees}{\raisebox{-.5pt}{\drawsquare{6.5}{0.4}}\hskip-0.4pt%
          \raisebox{-.5pt}{\drawsquare{6.5}{0.4}}\hskip-0.4pt%
          \raisebox{-.5pt}{\drawsquare{6.5}{0.4}}}
\newcommand{\Yasymm}{\raisebox{-3.5pt}{\drawsquare{6.5}{0.4}}\hskip-6.9pt%
        \raisebox{3pt}{\drawsquare{6.5}{0.4}}}

\newcommand{\Yadjoint}{\raisebox{-3.5pt}{\drawsquare{6.5}{0.4}}\hskip-6.9pt%
        \raisebox{3pt}{\drawsquare{6.5}{0.4}}\hskip-0.4pt
        \raisebox{3pt}{\drawsquare{6.5}{0.4}}}
\newcommand{\jref}[4]{{\it #1} {\bf #2}, #3 (#4)}

\newcommand{\NPB}[3]{\jref{Nucl.\ Phys.}{B#1}{#2}{#3}}

\newcommand{\PLB}[3]{\jref{Phys.\ Lett.}{#1B}{#2}{#3}}

\newcommand{\PRD}[3]{\jref{Phys.\ Rev.}{D#1}{#2}{#3}}

\begin{document}

\begin{titlepage}
\begin{center}
{\hbox to\hsize{ \hfill  DOE/ER/40561-329-INT97-00-173}}
{\hbox to\hsize{               \hfill  UW/PT-97-14}}

\bigskip
\bigskip
\bigskip
\vskip.2in

{\Large \bf Strongly Coupled Supersymmetry as  \goodbreak the  Possible Origin
of Flavor \footnote{Talk  presented at Rencontres de Moriond ``Electroweak
Interactions and Unified Theories'', Les Arcs France, March 15-22, 1997, on
work by D. B. Kaplan, F.  Lepeintre, and M. Schmaltz \cite{ourpaper}.}} \\

\bigskip
\bigskip
\bigskip
\vskip.2in
{\bf David B. Kaplan}\\

\vskip.2in

{ 
Institute for Nuclear Theory,   Box 351550

University of Washington,

Seattle, WA 98195-1550
}

\smallskip

{\tt dbkaplan@phys.washington.edu}

\vspace{1.5cm}
{\bf Abstract}\\
\end{center}

\bigskip

The existence of the electroweak hierarchy gives one strong reason to suspect
that nonperturbative physics lurks between the electroweak and the Planck
scales.  In this talk I speculate that these nonperturbative interactions might
also be behind the fascinating but obscure flavor structure we see at low
energies in the form of  fermion masses and mixing angles. A ``dual''
Froggatt-Nielsen mechanism is shown to explain how hierarchies can arise from
quark and lepton substructure.

\bigskip

\end{titlepage}

\section{New strong interactions and flavor}
\begin{figure}
\centerline{\epsfxsize=12 cm \epsfbox{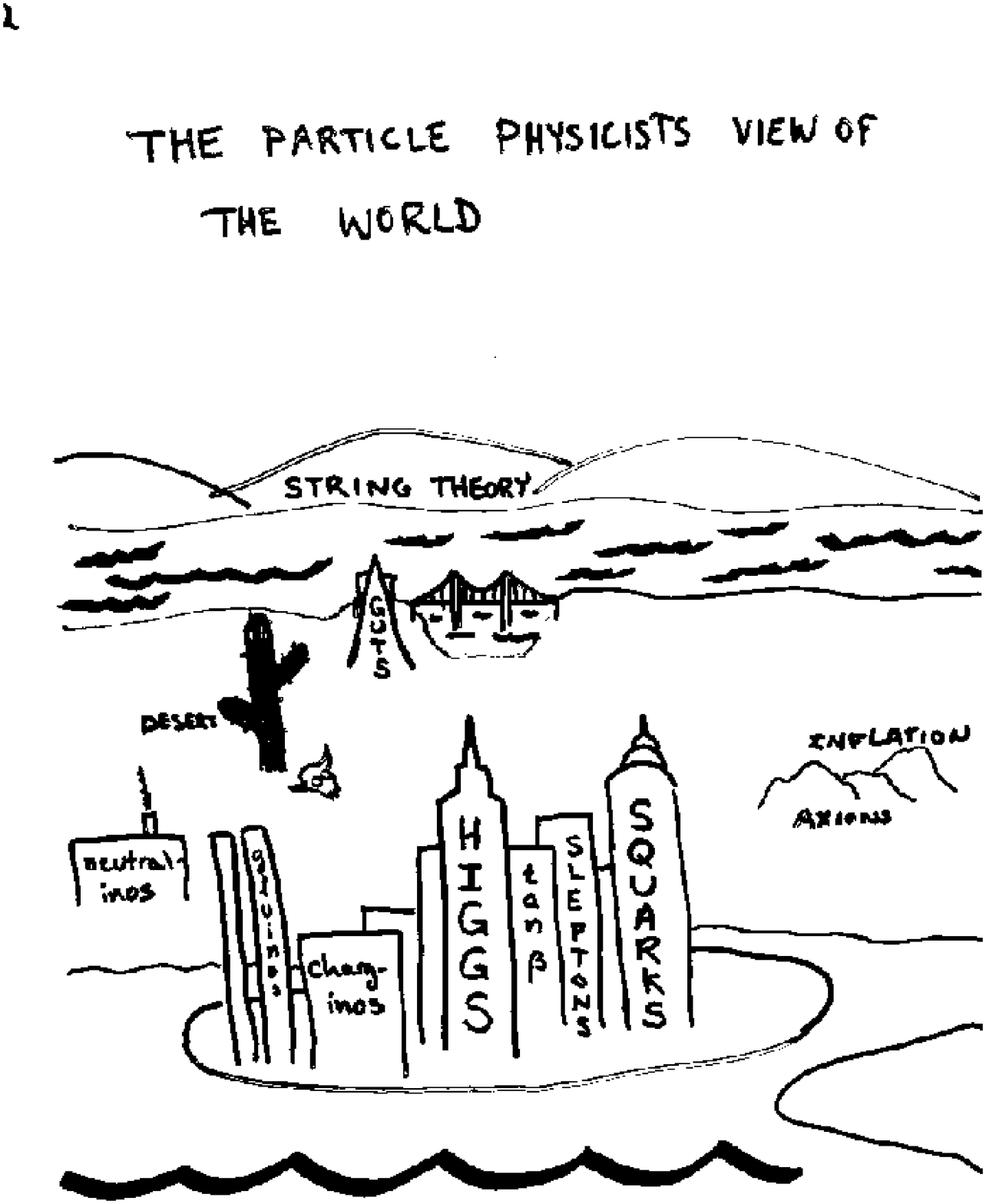}}
\vskip.2in
\end{figure}

A commonly held picture of physics beyond the standard model these days
consists of the minimal supersymmetric extension, a desert with perturbative
interactions up to the GUT scale, and just beyond, string theory.  Occasional
oddities populate the desert in various versions --- perhaps an axion, an
inflaton, or massive right-handed neutrinos.  The machinery required for
supersymmetry breaking is tastefully hidden behind a veil and called the hidden
sector.

That machinery is expected to entail new strong interactions.  The motivation
is that the large hierarchy between the Planck and electroweak scale looks very
much like the large hierarchy between the Planck and the QCD scales --- after
all, the QCD and electroweak scales only differ by two decades.  The latter is
due to nonperturbative physics, and results from the large value of
$e^{1/\alpha_s}$, where $\alpha_s$ is the QCD coupling at the Planck scale.
Similarly, it is plausible that the electroweak scale (or the SUSY breaking
scale, in a supersymmetric theory) arises from some new nonperturbative
physics.  This ought to be an exciting conclusion, since one might expect
nonperturbative physics to entail a rich phenomenology for experimentalists to
explore.  However, the success of precision electroweak predictions in the
standard model, the absence of observable flavor changing processes, the
stability of the proton, and the apparent perturbative unification of coupling
constants all serve to make new nonperturbative physics an unwelcome guest at
the party.  So it is usually relegated to the hidden sector where it won't
bother anyone.

In spite of the prevalent prejudice, it  seems to my collaborators and me worth
exploring whether  new strong interactions  at short distance scales might not
also be useful in explaining the outstanding question in particle physics ---
namely the origin of flavor. There are few hints as to why we have three
families, and  why fermions have the peculiar mass ratios and mixing angles we
observe.  We have seen in the hadronic spectrum that complicated structure can
arise from nonperturbative physics, and it seems natural to explore whether the
spectrum of the patterns seen in the Yukawa interactions of the standard model
might likewise be the result of strong dynamics.  Our motivation is in part
opportunistic --- with the recent advance in understanding strongly coupled
supersymmetric theory \cite{Seib,exactres}, one can now reliably discuss
theories with massless composite fermions, and so it is now at last possible to
explore an old idea that the distinctions between families are due to internal
substructure.
\section{The flavor problem}

Guessing at the dynamics underlying the quark and lepton masses and mixing
angles is particularly difficult since one needs more information than is
experimentally accessible. We would like to measure the 54 real parameters
characterizing the three (up quark, down quark and charged lepton), $3\times
3$, complex  Yukawa matrices of the standard model; probably there is some
natural basis for these matrices which exposes the underlying interactions that
give rise to the observed flavor structure.  However, out of those 54
parameters, all  we can measure at low energy are thirteen:  the three charged
lepton and six quark masses, as well as the three angles and one complex phase
in the CKM matrix (for simplicity I will ignore the possibility of neutrino
masses).
In the pure standard model, the
 other parameters are unphysical and can be eliminated by redefining the
quark and lepton fields.

In an extension of the standard model which has flavor physics at short
distances, many additional parameters of the Yukawa matrices are physical, and
are encoded in higher dimension operators, such as  dimension 6 four-fermion
interactions (including flavor changing neutral currents, proton decay, etc.).
When such operators are present in the theory, redefining the quark and lepton
fields does not eliminate the ``unphysical'' parts of the Yukawa matrices, as
they reappear in flavor changing (and possibly CP violating) higher dimensional
operators.  Unfortunately, such operators are experimentally inaccessible if
the scale of the new physics is above $\sim 1000$ TeV.

\section{Texture and broken flavor symmetry}
A popular recourse is to guess at what possible form the Yukawa matrices take
in the basis that is natural from the point of view of the short distance
flavor physics, and then to try to construct the dynamics that would give rise
to those Yukawa matrices.  The game is ``guess the texture'', where texture
refers to the pattern and hierarchies of entries that appear in the Yukawa
matrices.  For example, in ref. \cite{LNS}, a model is described in which the
Yukawa matrices for the up and down quarks look like
\beq
Y_{up} = \left(\matrix{
\epsilon_2^2 & 0 &\epsilon_2\cr
\epsilon_1\epsilon_2 &\epsilon_2&\epsilon_1\cr
\epsilon_2 &0&1\cr}\right) \qquad
Y_{down} = \left(\matrix{
\epsilon_2^2& \epsilon_1\epsilon_2 &\epsilon_1\epsilon_2\cr
\epsilon_1\epsilon_2 &\epsilon_1^2&\epsilon_1^2\cr
\epsilon_2 &\epsilon_1&\epsilon_1\cr}\right) \ ,
\eeq
with the predictions
\beq
\eqn{texture}
\vert V_{cb}\vert \sim {m_s\over m_b} \sim {m_b\over m_t} \sim \epsilon_1
\qquad
\vert V_{ub}\vert \sim {m_u\over m_c} \sim {m_c\over m_t} \sim \epsilon_2
\qquad
\vert V_{us}\vert \sim \sqrt{m_s\over m_b} \sim {\epsilon_2\over \epsilon_1}
\eeq
which looks reasonable with
$\epsilon_1 \sim {1\over 25}$, $\epsilon_2\sim {1\over 120}$.

In turn, the texture \refeq{texture} can be explained as arising from broken
approximate flavor symmetries. In the standard model, in the limit that the
Yukawa matrices vanish, there is a $U(3)^5$ chiral symmetry --- a $U(3)$ family
symmetry for each of the five different fermion representations ($q$, $\ell$,
$u^c$, $ d^c$, $e^c$).  With realistic Yukawa matrices, the only exact remnant
of this symmetry is $U(1)_B\times U(1)_L$, lepton and baryon number (ignoring
the electroweak anomaly).  However, a much larger subgroup of the $U(3)^5$
remains as an approximate symmetry, since some of the fermion Yukawa
interactions are so weak (\eg, the electron).  For example, if we approximated
the real world by setting all Yukawa couplings to zero except the top quark's,
the flavor symmetry would be $U(3)^3 \times U(2)^2 \times U(1)\subset U(3)^5$.

What is confusing when trying to guess short distance physics is that some
approximate flavor symmetries at low energy might also be  approximate
symmetries at high energy,  others  might be {\it accidental} symmetries that
are badly broken  at high energy. An example of the latter is baryon number in
$SU(5)$ --- it  is approximately conserved at low energy, even though
interactions at the GUT scale violate it badly. If one assumes that some
subgroup $H\subset U(3)^5$ is an approximate flavor symmetry at the scale of
flavor physics, and is broken by small parameters $\epsilon$, then the Yukawa
matrices are forced to have a certain texture built up out of powers of the
$\epsilon$ parameters,  according to how the various fermions transform.   In
the above example, the texture \refeq{texture} can be explained by an
approximate $U(1)\times U(1)$ flavor symmetry at high energy broken by the
parameters $\epsilon_{1,2}$, with appropriate charge assignments for the quarks
\cite{LNS}. (Note that in this example, the limit $\epsilon_{1,2}\to 0$  makes
the $U(1)^2$ symmetry exact at high energy, while the accidental  symmetry
group of the renormalizable interactions at low energy is the much larger
$U(3)^2\times U(2)\times U(1)$.)

The game becomes one of finding a fundamental approximate flavor symmetry $G$,
choosing the representations of both the fermions and  the small parameters(s)
$\epsilon$ which break $G$ explicitly, thus determining the Yukawa texture.
This idea was first discussed in detail by Frogatt and Nielsen \cite{FN}, who
broke the flavor symmetries spontaneously, so that the $\epsilon$ parameters
were proportional to various scalar vacuum expectation values. One then must
see if the model explains the observed masses and mixings, while being
consistent with FCNC constraints.  The drawback with this procedure is that
there is quite a lot of freedom in choosing symmetries and charges, and little
predictability.

\section{Families from compositeness}

The type of model mentioned above considers solutions to the flavor problem
which are perturbative.  In ref. \cite{ourpaper}, Fran\c{c}ois Lepeintre,
Martin Schmaltz and I propose how flavor structure could arise from new strong
interactions.  The basic idea is that quarks and leptons, and possibly the
Higgs, might be composite at short distance scales, with Yukawa interactions
being generated by interactions among the constituents.  Families are
distinguished by different constituents, and therefore have different strength
couplings to the Higgs.  Thanks to recent advances in understanding strongly
coupled SUSY theories, it is possible to discuss such theories with some
confidence.

 A possible paradigm is found in nuclear physics.
Consider, for example, the three isotopes of hydrogen: they
each have the same chemistry  yet have dramatically
different masses --- a fact simply  understood once it is realized that the
nucleus is composite, and that the three isotopes each contain a single proton
but varying numbers of neutrons.  Similarly, quarks and leptons could be
bound states of both charged and neutral constituents, with different numbers
or types of neutral constituents for the different generations.  The nature of
these composites will be determined by the underlying strong interactions, and
the interactions of the ``neutrons'' will largely determine the flavor
structure observed at low energies.

One consequence of this picture is that the number of families can be
determined by the properties of the strong dynamics, much as the number of
isotopes of hydrogen is determined by the strong interactions.  Another
consequence is that the approximate flavor symmetry at short distances will be
a product of $U(1)$'s which essentially count constituents. Continuous
nonabelian symmetries won't appear because particles with different numbers of
consituents will be dissimilar when regarded closely, even if they have similar
standard model properties.  Texture in the Yukawa interactions can easily arise
since, with different numbers of constituents in the different families, the
interactions giving rise to Yukawa couplings will have different dimensions ---
and thus different strengths --- depending on which families participate.

An interesting example of a composite model that predicts three families is an
$Sp(6)$
gauge theory with 6 fundamentals
$Q$ and an antisymmetric tensor $A$ \cite{Cho,oursp}.  The theory
has an $SU(6)\times U(1)$ global symmetry, as well as an $R$ symmetry.  The
confined description involves the  $Sp(6)$ neutral fields
\beq
T_m &=& {\rm Tr} A^m, \ m=2,3\nonumber\\
M_n &=& Q A^n Q, \ n=0,1,2\ ,
\eeq
where $Sp(6)$ indices are contracted with the appropriate metric, which can be
taken to be $J = i\sigma_2\times I_3$. The number  of $M$-particle families
equals
$3$ because of properties of $Sp(6)$ representations.  The quantum
numbers of the fields
are:

\beq
 \begin{array}{c|c|crc}
    & Sp(6) & SU(6) & U(1) & U(1)_R \\[.1in] \hline
&&&&\\[-.1in]
  A & \Yasymm   & 1     & -3   & 0 \\
  Q & \Yfund    & \Yfund& 2& \frac{1}{3} \\[.1in] \hline \hline&&&&\\[-.1in]
  T_m    & & 1       & -3m  & 0 \\
  M_n   &  & \Yasymm & 4 -3 n & \frac{2}{3}
 \end{array}
\eqn{sp2n}
\eeq
I have indicated  symmetry representations by   Young tableaux in this table:
$\Yfund$ is the defining representation, while $\Yasymm$ denotes an
antisymmetric tensor.
This model has a number of desirable features. If weak gauge interactions are
embedded in the $SU(6)$ symmetry of this model, then there is a replication of
``families'' of $3$ fields~\cite{oursp}. Furthermore,
in spite of having $3$ families, the  family symmetry of the model is not
$U(3)$,  but only
$U(1)$. Family replication arises because the $A$ field only carries this
global $U(1)$ charge, and so the SM gauge charges of a composite particle are
independent of the number of $A$ fields it contains.  Breaking this $U(1)$
flavor symmetry will allow us to generate flavor in a manner analogous to the
Froggatt-Nielsen mechanism. The model realizes the isotope paradigm of the
introduction, with  the $(QQ)$ and $A$ fields  playing the roles  of the proton
and neutron respectively.

\section{Texture from compositeness}

In order to create a realistic model based on the family replication example
above, one needs more than a single $Sp(6)$ group, as a family cannot fit in
the antisymmetric tensor of $SU(6)$.  A possible extension is to use an
$Sp(6)^3$ gauge group, where one $Sp(6)$ binds together the left-handed quarks,
the second  confines the right-handed quarks, and the third produces composite
Higgs fields.  The three groups could be made to talk to each other by having
three massive fields $V_{1,2,3}$, each of  which transforms as a fundamental
under two of the $Sp(6)$ groups at a time.  On can introduce $VQQ'$, $VVA$ and
$VVV$ type interactions in the superpotential.  When the $V$ fields are
integrated out of the theory at their mass scale (assumed to be above the
confining scale), they produce multifermion interactions; after confinement,
these lead to Yukawa interactions among the composite fields (Fig.~1).  If the
$VVA$ coupling is small, then every time an additional $A$ field is included in
the graph, there is a suppression factor.  Since the number of $A$ constituents
distinguishes families, the theory provides a new mechanism for generating
texture.  Graphs of the sort pictured in Fig.~1 can give rise to Yukawa
coupling matrices of the form
\begin{figure}[t]
\centerline{\epsfxsize=2 in \epsfbox{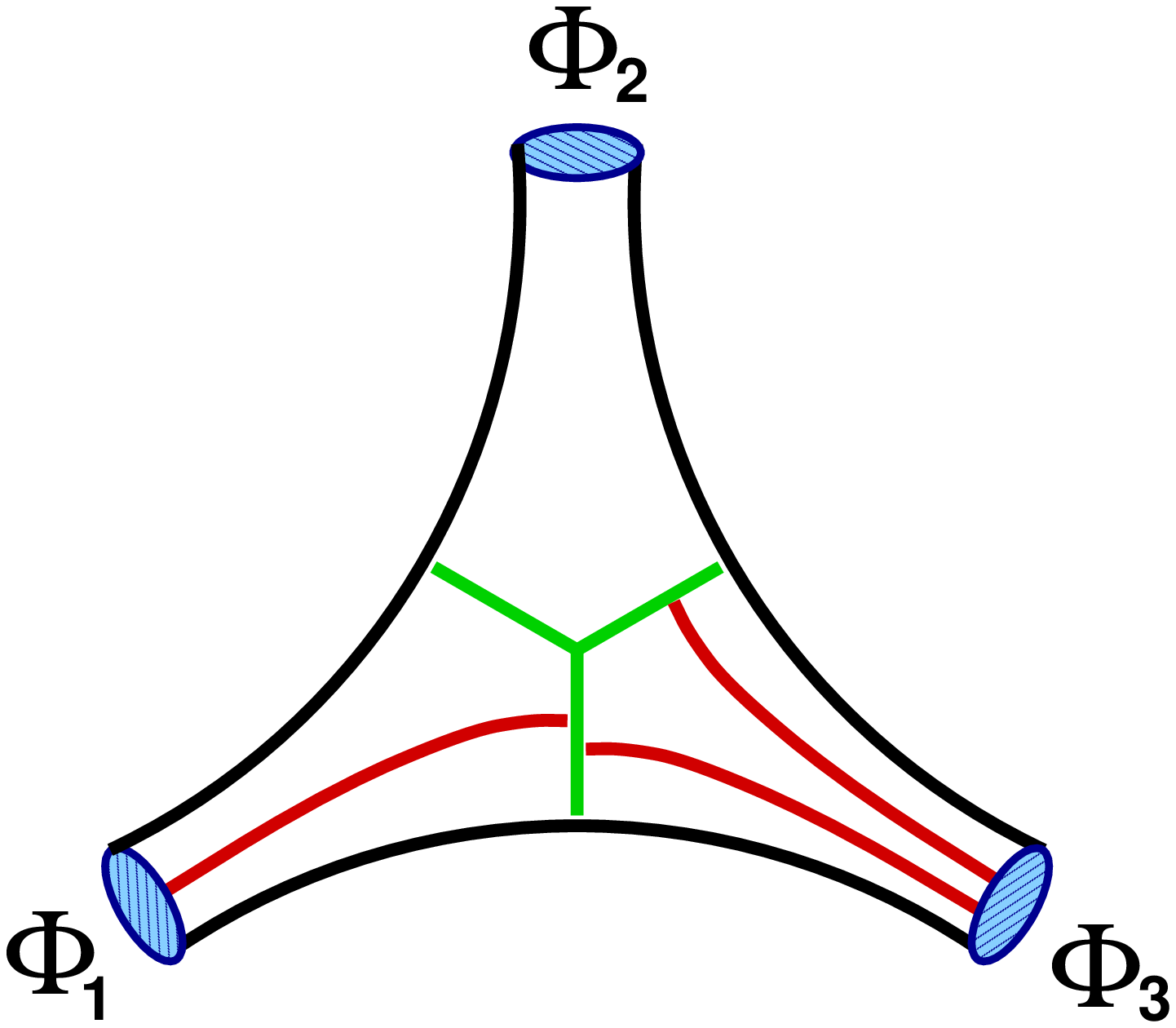}}
\vskip.2in
\noindent
Fig 1. {\it Contribution to the effective superpotential
from integrating out the massive $V$  fields. Below the confinement
scale,
these contributions become trilinear interactions (Yukawa interactions) between
composite fields.  The strength of the effective Yukawa interaction is less for
composite fields with more of the $SU(3)\times SU(2)\times U(1)$ neutral $A$
constituents, which attach to the $V$ propagators.}
\vskip .2in
\end{figure}

\beq
\eqn{yzero}
 Y\propto \left(\matrix{
19 \epsilon^4 &9 \epsilon^3 &3 \epsilon^2 \cr
9 \epsilon^3  &5 \epsilon^2 & 2\epsilon\cr
3 \epsilon^2&2\epsilon & 1\cr
}\right)
\eqn{toyyuk}
\eeq
where $\epsilon = g_{VVA} \Lambda/M$, with $g_{VVA}$ being the $VVA$ coupling,
$M$ being the mass of  the $V$ fields, and $\Lambda$ being the confinement
scale of the $Sp(6)$ groups.  The peculiar numbers appearing in \Eq{yzero}
arise as combinatoric factors. The eigenvalues of the matrix $Y$ are
approximately \{1,$\epsilon^2$,$\epsilon^4$\}, and could explain the hierarchy
observed in up quark masses if $\epsilon\sim 1/20$, for example. This mechanism
is in a sense dual to the Frogatt-Nielsen proposal, as the $A$ fields are
confined, as opposed to getting vacuum expectation values.

This example shows quantitatively how texture can arise from compositeness.
The three families are distinguished by the number of $A$ constituents, and the
$U(1)$ symmetries carried by the $A$ fields would prohibit masses and mixings
among families.  However the $VVA$ and $VVV$ couplings violates these $U(1)$'s
explicitly, and lead to masses and mixing. Particles with many $A$ constituents
require many powers of $g_{VVA}$ in their masses, and are hence quite light.
This mechanism for generating texture is more constrained than the
Frogatt-Nielsen mechanism since the structure and  of the light composite
fermions are fixed by the strong dynamics and can't be tweaked to one's desire.
In ref. \cite{ourpaper} more complicated and realistic examples are given.

\section{What should we look for?}

I began this talk motivating the existence of new strong interactions by the
existence of the electroweak hierarchy, and then proceeded to ignore SUSY
(electroweak) symmetry breaking.  Putting flavor and SUSY breaking together in
a single strongly interacting theory is an interesting and challenging
enterprise.  It is hard to even figure out where to begin.

One possible starting point is to ask whether the quantum numbers of the
standard model particles suggest any particular compositeness structure.  For
inspiration, consider the quark model and the way it explains the baryon
spectrum.  If one classifies the baryon octet and decuplet first under $SU(2)$
isospin, then $SU(3)$, then finally $SU(6)$, one finds the following
representations (again I use Young tableaux, as they provide a particularly
suggestive way to visualize group representations):
\beq
\begin{array}{cclcc}
 N  \oplus\Lambda \oplus \Sigma  \oplus \Xi \oplus\Delta \oplus \Sigma^*
\oplus\Xi^*\oplus  \Omega
&=& \Yfund  \oplus 1 \oplus \Ysymm \oplus \Yfund \oplus\Ythrees \oplus \Ysymm
\oplus\Yfund \oplus  1&  &SU(2)\cr
 &\rightarrow& \Yadjoint \oplus\Ythrees&  &SU(3)\\
 &\rightarrow& \Ythrees&  &SU(6)
\end{array}
\eeq
These representations do not provide striking evidence for compositeness at the
level of $SU(2)$;  under  $SU(3)$ one sees a threefold structure, but the
symmetry properties are obscure;  however, classification under $SU(6)$
suggests immediately that the baryons are a  bound state of three constituents
totally symmetric under spin/flavor.

The analogous procedure for the standard model particles is to look at their
charges when unified into larger and larger groups, then gaze at the relevant
Young tableaux and see if any particular substructure suggests itself.
A clue that this is the right procedure to follow  is the apparent perturbative
unification of the standard model coupling constants in a manner consistent
with $SU(5)$,  groups  which contain $SU(5)$, such as  $SO(10)$ and  $E_6$, and
 $SU(3)\times SU(3)\times SU(3)$.   $SU(5)$ representations for a family
($\mybar\Yfund \oplus \Yasymm$) do not readily suggest substructure, unless one
says that the $\mybar\Yfund$ is fundamental and the $\Yasymm$ is a bound state
of a pair of constituents \cite{strassler,msan}.  Under $SO(10)$, a family is a
spinor, and if composite,  at least one of the constituents would likewise have
to be a spinor. So $SO(10)$  doesn't readily suggest compositeness, and neither
does  $E_6$ for similar reasons.

 In an $SU(3)\times SU(3)\times SU(3)$ GUT,  families do have structure that
suggests compositeness:
\beq
{\rm one\ family} =(\Yfund,\mybar \Yfund,1) \oplus (1,\Yfund,\mybar \Yfund)
\oplus (\mybar \Yfund,1,\Yfund)
\eeq
which has a natural explanation if each particle in the standard model is a
boundstate of two constituents, a fermion and a boson.  This is the structure
explored in ref. \cite{ourpaper}, but none of the realistic examples there are
consistent with unification.  It must be admitted that maintaining $SU(3)\times
SU(2)\times U(1)$ coupling constant unification in a theory of composite quarks
and leptons is not easy, although  possible  \cite{strassler}).

\section{Conclusions}

Perhaps for a physicist today  to try to understand the origin of quark and
lepton masses with our limited experimental information about flavor is
analogous to asking Mendel to try to deduce the structure of DNA from his
experiments on breeding peas.  Nevertheless, it is the outstanding question for
particle theorists to answer, and attempts to do so may eventually lead to a
solution, if not a sudden epiphany.  The work described here uncovers a new
mechanism that is capable of generating the observed spectrum we observe. One
of the virtues of our work is that the models presented are renormalizable
field theories, so that none of the required dynamics is  hidden. By analyzing
quantitatively  how mass hierarchies can arise from internal structure for
quarks and leptons, a new paradigm is offered which is hoped to contain a germ
of truth.

Putting aside the usual apologies and lies that are the common refuge of
theorists, I,
like many of my experimental colleagues at this conference, must conclude that
my collaborators and I  have looked for something radically new --- 
in this case a composite model
that explains flavor, unification, and SUSY breaking ---  but that we have not
found it yet at the 95\% confidence level.  Nevertheless, we have seen
promising signs of something new on the horizon, and have high hopes for the
next run.

\section{Acknowledgements}
I thank M. Schmaltz for going over this manuscript.
This work is  supported in part by DOE grant
DOE-ER-40561, and NSF Presidential Young Investigator award
PHY-9057135.


\end{document}